\documentclass[aps,twocolumn,prb,superscriptaddress,showpacs,showkeys]{revtex4-1}
\usepackage{graphicx}

\begin{document}

\title{Edge and Interfacial States in a 2D Topological Insulator:\\Bi(111) Bilayer on Bi$_{2}$Te$_{2}$Se}

\author{Sung Hwan Kim}
\affiliation{Center for Artificial Low Dimensional Electronic Systems, Institute for Basic Science (IBS), 77 Cheongam-Ro, Pohang 790-784, Korea}
\affiliation{Department of Physics, Pohang University of Science and Techonlogy, 77 Cheongam-Ro, Pohang 790-784, Korea}

\author{Kyung-Hwan Jin}
\affiliation{Department of Physics, Pohang University of Science and Techonlogy, 77 Cheongam-Ro, Pohang 790-784, Korea}

\author{Joonbum Park}
\affiliation{Department of Physics, Pohang University of Science and Techonlogy, 77 Cheongam-Ro, Pohang 790-784, Korea}

\author{Jun Sung Kim}
\affiliation{Department of Physics, Pohang University of Science and Techonlogy, 77 Cheongam-Ro, Pohang 790-784, Korea}

\author{Seung-Hoon Jhi}
\email{jhish@postech.ac.kr}
\affiliation{Department of Physics, Pohang University of Science and Techonlogy, 77 Cheongam-Ro, Pohang 790-784, Korea}

\author{Tae-Hwan Kim}
\affiliation{Center for Artificial Low Dimensional Electronic Systems, Institute for Basic Science (IBS), 77 Cheongam-Ro, Pohang 790-784, Korea}
\affiliation{Department of Physics, Pohang University of Science and Techonlogy, 77 Cheongam-Ro, Pohang 790-784, Korea}

\author{Han Woong Yeom}
\email{yeom@postech.ac.kr}
\affiliation{Center for Artificial Low Dimensional Electronic Systems, Institute for Basic Science (IBS), 77 Cheongam-Ro, Pohang 790-784, Korea}
\affiliation{Department of Physics, Pohang University of Science and Techonlogy, 77 Cheongam-Ro, Pohang 790-784, Korea}

\date{\today}

% Abstract
\begin{abstract}
The electronic states of a single Bi(111) bilayer and its edges, suggested as a two dimensional topological insulator, are investigated by scanning tunneling spectroscopy (STS) and first-principles calculations. Well-ordered bilayer films and islands with zigzag edges are grown epitaxially on a cleaved Bi$_{2}$Te$_{2}$Se crystal. The calculation shows that the band gap of the Bi bilayer closes with a formation of a new but small hybridization gap due to the strong interaction between Bi and Bi$_{2}$Te$_{2}$Se. Nevertheless, the topological nature of the Bi bilayer and the topological edge state are preserved only with an energy shift. The edge-enhanced local density of states are identified and visualized clearly by STS in good agreement with the calculation. This can be the sign of the topological edge state, which corresponds to the quantum spin Hall state. The interfacial state between Bi and Bi$_{2}$Te$_{2}$Se is also identified inside the band gap region. This state also exhibits the edge modulation, which was previously interpreted as the evidence of the topological edge state [F.~Yang et al., Phys.~Rev.~Lett.~\textbf{109}, 016801 (2012)]. 
\end{abstract}

\maketitle

\section{Introduction} 
In the last few years, the study of topological degrees of freedom has become one of the most actively pursued  topics in condensed matter physics. In terms of materials, this stream of researches has been triggered by the discovery of the two dimensional (2D) topological insulator (TI), the HgTe/CdTe quantum well,\cite{bernevig_quantum_2006, konig_quantum_2007} and followed by the outbursting discovery of various 3D TI crystals starting with Bi tellurides (selenides).\cite{hsieh_topological_2008, chen_experimental_2009, hasan_colloquium:_2010, kuroda_experimental_2010, souma_topological_2012, al-sawai_topological_2010, liu_metallic_2011, sun_new_2010} These materials have non-degenerated spin-polarized Dirac fermion bands on their surfaces,\cite{zhang_topological_2009} which are protected by the time-reversal symmetry. In the 2D case, these bands correspond to quantum spin Hall (QSH) states, which bear important potentials for spintronic applications and the quantum computing.\cite{murakami_dissipationless_2003, sinova_universal_2004}

In sharp contrast to various different types of 3D TI materials discovered, 2D material systems to realize the QSH state have been extremely limited to HgTe/CdTe and InAs/GaSb.\cite{bernevig_quantum_2006, konig_quantum_2007,InAs_theory, InAs_expt} Moreover, while the existence of metallic edge states was confirmed directly by spin- and angle-resolved photoemission spectroscopy (ARPES) and scanning tunneling spectroscopy (STS) for 3D TI's, no such study is available for 2D TI's; the QSH states of HgTe and InAs were revealed only by the transport measurements.\cite{konig_quantum_2007, InAs_expt} In this respect, recent progress in the growth and characterization of one bilayer (BL) Bi(111) films, which were theoretically suggested as a 2D TI,\cite{murakami_quantum_2006,wada_localized_2011} deserves particular attention.\cite{sakamoto_spectroscopic_2010, hirahara_interfacing_2011, liu_stable_2011, jnawali_manipulation_2012} So far, the ARPES and transport measurements on this film could only probe the bulk 2D property without providing any information on the edge electronic state,\cite{hirahara_interfacing_2011,jnawali_manipulation_2012} which is the essential feature of 2D TI, due to the extremely low density of edges in the film. On the other hand, a very recent scanning tunneling microscopy (STM) and STS study claimed that the edge-localized density of states (DOS) of a Bi(111) BL island were resolved.\cite{yang_spatial_2012} However, the DOS enhancement itself was marginal and the corresponding energy was not clearly separated from other spectral features nor consistent with the calculation based on a freestanding Bi layer. More importantly, a strong hybridization of the electronic bands of the Bi BL and the substrate, Bi$_{2}$Te$_{3}$, was observed in both ARPES and first-principles calculations to drive the BL apparently \textit{metallic}.\cite{hirahara_interfacing_2011,liu_stable_2011,yang_spatial_2012,miao_2013} Therefore, \textit{the TI nature of the Bi BL film grown on Bi$_{2}$Te$_{3}$ is not clear} at present.

In this paper, through STM, STS, and ARPES measurements and extensive first-principles calculations, we investigated the electronic states of Bi(111) one BL islands with well-defined zigzag edges grown on fresh cleaved surfaces of Bi$_{2}$Te$_{2}$Se. The strong interaction between the Bi layer and the substrate is unraveled. The calculation reveals that the topological edge state (TES) of the Bi BL is preserved even after the the strong hybridization with the substrate, which replaces the band gap of the pristine Bi BL with a small hybridization gap. This gap, playing an important role to make the epitaxial Bi(111) BL a 2D TI, is confirmed by the STS measurements. More importantly, the edge electronic state was clearly distinguished from other spectral feature of the Bi(111) BL including the characteristic interfacial states, in good quantitative agreement with first-principles calculations for a Bi(111) BL film and BL nanoribbons on top of Bi$_{2}$Te$_{2}$Se. We suggest that the edge enhanced STS feature observed previously could be due to the edge modulation of the interfacial state.\cite{yang_spatial_2012} This work makes clear the 2D TI nature of a Bi(111) BL on a \textit{strongly interacting substrate} of a 3D TI Bi$_{2}$Te$_{3}$Se and opens up the possibility for microscopic studies of a quantum spin Hall channel. The present system is also unique due to the coexistence of the 2D and 3D topological insulators in proximity although its implication is largely unknown.

% Figure1
\begin{figure}[!pb]
\includegraphics{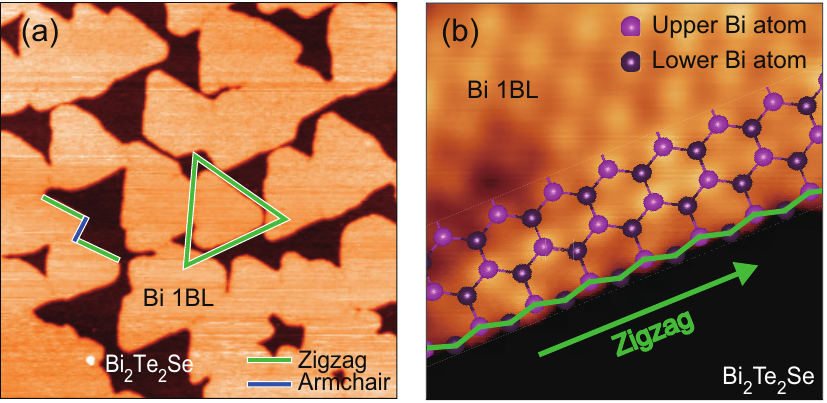}
\caption{(a) STM topography of Bi(111) BL islands on Bi$_{2}$Te$_{2}$Se (V$_{\textrm{s}}=1~$V, I$_{\textrm{t}}=100~$pA, 250~nm $\times$ 250~nm). (b) High resolution topography of a Bi(111) BL island and its edge. The schematic of the atomic structure of a Bi(111) BL is shown together. The zigzag and armchair edges are indicated by green and blue lines in (a) and (b).}
\end{figure}

\section{Experimental and theoretical methods}
We used Bi$_{2}$Te$_{2}$Se single crystals as substrates, which were grown using the self-flux method with a stoichiometry of chunks (Bi:Te:Se$=$2:1.95:1.05).\cite{ren_large_2010} Experiments were performed in ultrahigh vacuum better than 5$\times$10$^{-11}$ Torr, using a commercial low-temperature STM (Unisoku, Japan). Bi$_{2}$Te$_{2}$Se single crystals were cleaved in UHV, and well ordered surfaces were confirmed by atomically resolved STM measurements. Our STS ($dI/dV$) curves for the cleaved surface of Bi$_{2}$Te$_{2}$Se are fully consistent with the previous work.\cite{alpichshev_stm_2010} Ultrathin Bi(111) films were grown at room temperature by evaporating Bi from a Knudsen cell and samples were cooled down to $\sim$78 K for STM and STS measurements. Topography imaging was performed at several sample biases and currents, mainly 1 V and 100 pA. STS spectra were acquired using the lock-in technique with a bias-voltage modulation of 1 kHz at 10--30 mV$_{\textrm{rms}}$ and a constant tunneling current of 0.8--1nA. The energy band positions of both the substrate and the Bi(111) films were determined by ARPES measurements using Xe discharge radiation (8.4 eV) and a high performance hemispherical electron energy analyzer (Gamma-Data, Sweden). 

% Computational Method
First-principles calculations were carried out in the plane-wave basis within the generalized gradient approximation for the exchange-correlation functional.\cite{Kresse_efficient_1996, Perdew_generalized_1996} A cutoff energy of 400 eV was used for the plane-wave expansion. A $k$-point mesh of 11$\times$11$\times$1 was used for sampling the Brillouin zone. The Bi(111)/Bi$_{2}$Te$_{2}$Se structure was simulated in a supercell with one BL Bi(111) on one surface of a Bi$_{2}$Te$_{2}$Se slab with a thickness of six quintuple layers (QLs) and a 2.0 nm thick vacuum layer. We used the experimental lattice constant of Bi$_{2}$Te$_{2}$Se, 0.436 nm.\cite{Sokolov_chemical_2004} The Bi(111) and three Bi$_{2}$Te$_{2}$Se QLs were relaxed fully. In order to investigate edge states, we carried out similar calculations for Bi(111) BL nanoribbons (BNRs) of different widths with zigzag edges on top of a three-QL-thick Bi$_{2}$Te$_{2}$Se slab. The thickness is not thick enough to secure the topological surface state of  Bi$_{2}$Te$_{2}$Se but affects little the electronic structure of the Bi(111) BL on top and its interaction with the substrate as confirmed by the calculations with a thicker film. While there is no previous calculation for an epitaxial BNR so far to be compared with ours to the best of our knowledge, our calculations for an epitaxial Bi(111) BL agree well with the previous works.\cite{wada_localized_2011, yang_spatial_2012}

\section{Results and discussions} 
The STM image in Fig.~1(a) shows that the Bi layer grows as triangular islands.\cite{jnawali_nucleation_2009} At certain coverage, these islands form a percolation network and the second layer starts to nucleate before the first layer completes. The height and the lateral unit cell of the islands were measured as $\sim$0.6 and $\sim$0.45 nm, respectively, corresponding well to those of a single Bi(111) BL. Individual atoms along islands edges were clearly resolved for the first time [Fig.~1(b)], which unambiguously indicate that the edges are mostly zigzag edges. Short armchair edges also appear as the minority case, one of which is marked as a blue line in Fig.~1(a).

In the middle of Bi(111) BL islands, we obtained STS ($dI/dV$) data with rich spectroscopic features [see the blue line in Fig.~2(b)]. In particular, inside the band gap region of the Bi(111) BL ($E_{gap, Bi}$) and the Bi$_{2}$Te$_{2}$Se ($E_{gap, BTS}$) substrate, we find three very prominent peaks ($\alpha$,~$\beta$~and~$\gamma$) at about $-$0.14, $+$0.03 and $+$0.4~eV, respectively (dashed lines).

% Figure2
\begin{figure}[!pt]
\includegraphics{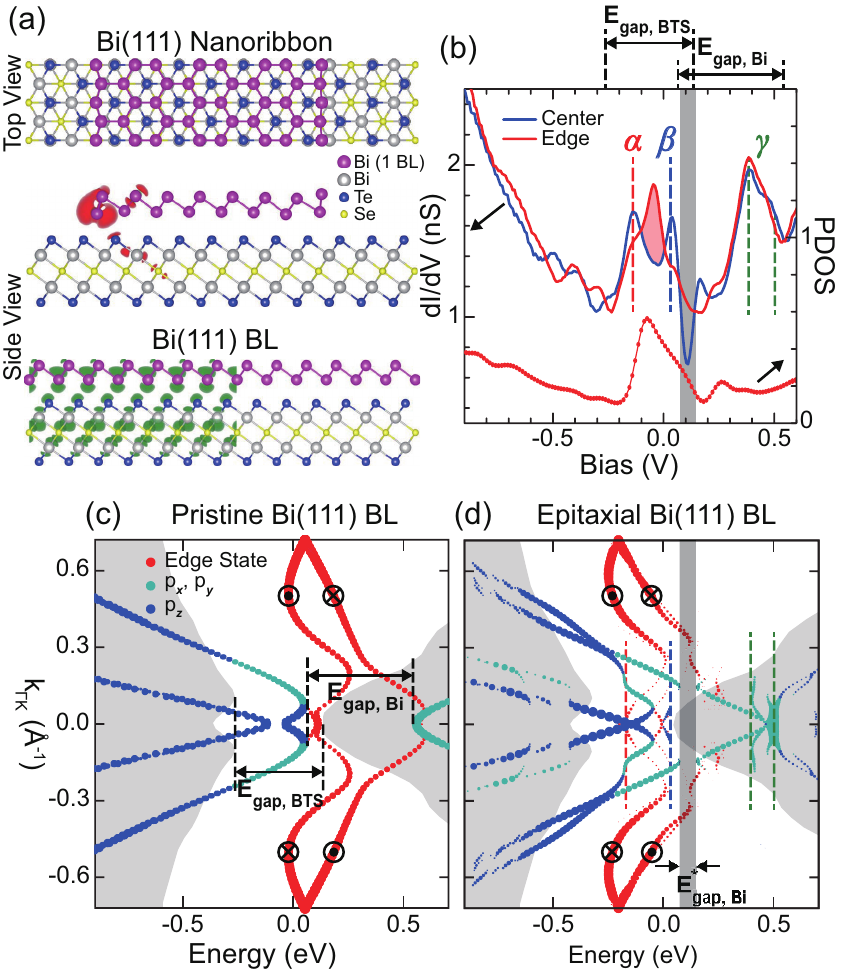}
\caption{(a) Atomic structures of a single Bi(111) BL (bottom panel) and a zigzag-edged single Bi(111) BL nanoribbon (BNR, top and middle panels) on top of Bi$_{2}$Te$_{2}$Se. Only part of substrate QLs are shown for clarity. (b) Calculated partial DOS originating from the BNR edge states together with corresponding STS ($dI/dV$) curves obtained at the center (blue) and the edge (red) of a Bi(111) BL. Calculated band structure along the $\Gamma$--K direction (c) for a free standing pristine Bi(111) BL and (d) for the epitaxial BL on the structure depicted in (a); the bands in blue and green originate from the Bi(111) BL and the substrate bands are colored in gray without showing the details for clarity. The orbital characteristics of the bands from Bi(111) BL are distinguished in color; green for $p_x$ ($p_y$) and blue for $p_z$. The bands in red depict the edge states of BNR (c) in the free-standing geometry and (d) on the substrate taken from Fig.~5. The spin orientations are marked in the edge bands by the heads and tails of arrows. The electron distributions for the edge states of BNR (red) and the interfacial state $\gamma$ at the bottom of the conduction band of Bi(111) BL (green) are also shown in (a).}
\end{figure}

% Figure3
\begin{figure}[!ptb]
\includegraphics{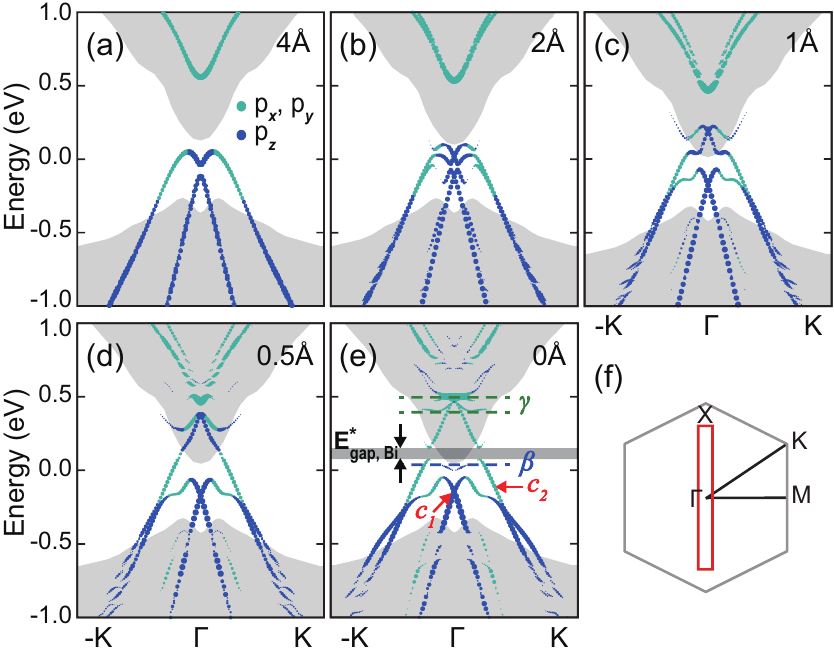}
\caption{(a)--(e) Calculated band structures of a single Bi(111) BL on top of Bi$_{2}$Te$_{2}$Se at a distance of 4.0, 2.0, 1.0, 0.5, and 0{\AA} from the equilibrium distance, respectively. The states originating from the Bi(111) BL are marked by blue (the $p_z$ orbital) and green ($p_x$ and $p_y$ orbitals) dots. The bands of the Bi$_{2}$Te$_{2}$Se substrate are located within the gray regions. (f) The hexagon represents the first Brillouin zone (BZ) of a Bi(111) BL on top of Bi$_{2}$Te$_{2}$Se and the red rectangle that of a Bi(111) nanoribbon on top of Bi$_{2}$Te$_{2}$Se used in our calculations.}
\end{figure} 

Our first-principles calculations reproduce these features clearly. As shown in Figs.~2(c), 2(d) and 3, the degenerate valence bands and conduction bands of Bi(111) BL are split due to the inversion symmetry breaking by the interaction with the substrate. We observe that one valence band splits off significantly and disperses sharply up in energy to the conduction band minimum (CBM) at $+$0.5 eV. There, the original band gap of the Bi(111) BL is closed and the hybridization with the CBM of the substrate leads to a small gap just above the Fermi energy ($E^*_{gap, Bi}$) [the gray bar in Fig.~2(d) and Fig.~3(e)], which is very obvious in the STS spectrum of Fig.~2(b). This unexpected \textit{hybridization} gap is, in fact, found to be very important for the TI nature as explained below. The occupied part of these band structure, in particular, the metallic band crossing the Fermi level, is consistent with the previous ARPES measurements as well as in our own work.\cite{hirahara_interfacing_2011,yang_spatial_2012, miao_2013} The above-mentioned spectral features in STS [Fig.~2(b)] are found to be due to large DOS at the flat parts of the bands as indicated by the dashed lines; those near the Rashba band-crossing in the valence band ($\alpha$), around the hybridization gap near the conduction band minimum of Bi$_{2}$Te$_{2}$Se ($\beta$), and near the conduction band edge of the pristine Bi(111) BL ($\gamma$). As shown in the charge density distribution in Fig.~4, $\alpha$ is mainly due to the valence bands of Bi(111) BL. On the other hand, $\beta$ and $\gamma$ originate from the conduction band minimum of Bi$_{2}$Te$_{2}$Se and the hybridized state between the Bi(111) BL and the substrate, respectively. We thus call $\gamma$ the interfacial state, which is the characteristic feature of the epitaxial Bi(111) BL [the green bulbs in Figs.~2(a) and 4(d)]. 

The interaction with the substrate changes the band structure so significantly that the topological nature and the edge state of Bi(111)/Bi$_{2}$Te$_{2}$Se cannot be analyzed using the bands of the floating Bi(111) BL as done in the previous works.\cite{hirahara_interfacing_2011,yang_spatial_2012} Thus, we calculated the edge electronic structure in the \textit{epitaxial geometry}. We used zigzag-edged Bi(111) BNRs on top of Three-QL-thick Bi$_{2}$Te$_{2}$Se film as shown in Figs.~2(a) and 5 with a width of 8--11 Bi zigzag chains. The thickness of 3~QLs is needed to reduce the computation load, which is not thick enough to secure the topological surface state of Bi$_{2}$Te$_{2}$Se.  We, however, checked that this thickness reduction affects little the electronic structure of the Bi(111) BL and its edge state. Our calculations reproduce clearly the edge-state bands, whose dispersion is surprisingly consistent with that of a free-standing Bi(111) BL qualitatively.\cite{murakami_quantum_2006, wada_localized_2011} The overall band dispersion, the Dirac cone at the zone boundary and the helical spin texture of the edge states are preserved while its energy shifts substantially to a higher binding energy by about 0.3 eV. This shift reflects the formation of the new hybridization gap; the CBM shifts from $+$0.5 to $+$0.15 eV and the edge state emerges from this new CBM. The energy shift is crucial to locate the edge state in the experiment as discussed below. The preservation of the edge states is guaranteed since the band inversion, the essential characteristics of a TI, of the Bi(111) BL survives in spite of the strong interaction with the substrate as depicted in Figs.~2(c), 2(d) and 3. Based on the band structure of the Bi(111) BL and its edge, we confirm that the epitaxial Bi(111) BL on Bi$_{2}$Te$_{2}$Se is still a 2D TI with a small band gap (the hybridization gap, $E^*_{gap, Bi}$) having a single TES crossing this gap. The fine splittings of TES shown in Fig.~2(d) are artifacts due to the finite width of the nanoribbon used in the calculation, which become less obvious in the calculation for wider ribbons [Fig.~5].

% Figure4
\begin{figure}[!pt]
\includegraphics{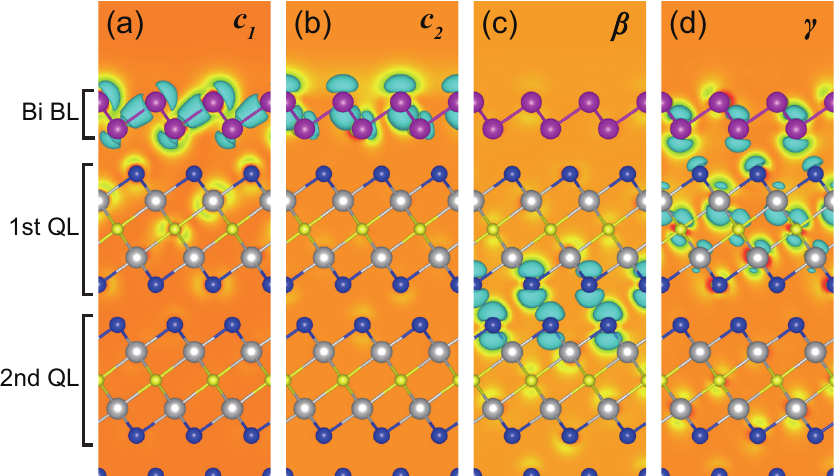}
\caption{(a)--(d) The charge density plots for the valence bands of a single Bi(111) BL on Bi$_{2}$Te$_{2}$Se in the optimized geometry at the $k$ points of $\beta$, $\gamma$, and $\alpha$ ($c_1$ and $c_2$) in Fig.~3(e). Green blobs denote the iso-surface of the charge density (1.5 $\times$ 10$^{-5}$/a$_{0}^{3}$, a$_{0}$ being the Bohr radius). Purple balls represent atoms in the Bi BL; gray, blue, and yellow balls Bi, Te, and Se atoms, respectively, in the substrate.}
\end{figure}

% Figure5
\begin{figure}[!pb]
\includegraphics{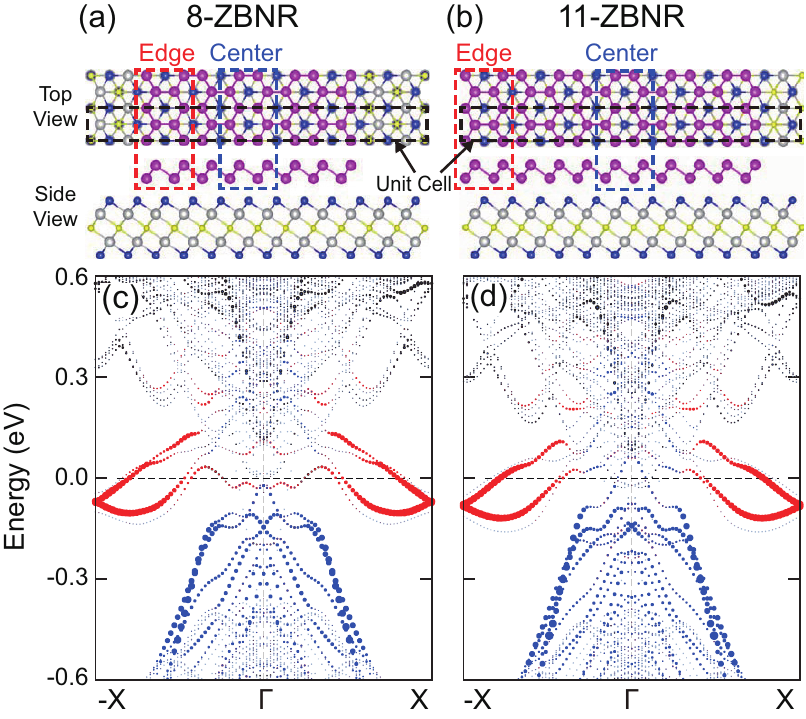}
\caption{(a) and (b) Schematics of the atomic structures of BNR's with zigzag edges on both sides. Two BNRs with widths of 8 and 11 Bi zigzag chains, 8-ZBNR and 11-ZBNR respectively, are shown. Calculated band structures of (c) [(d)] the 8-ZBNR [11-ZBNR] on Bi$_{2}$Te$_{2}$Se 3 QLs based on the atomic structure in (a) [(b)]. The states originating from edge (center) atoms are marked by the red (blue) dots. The band width of the edge states are $\sim$0.26~eV, which increases to $\sim$0.38~eV after the edge relaxation [Fig.~2(d)].}
\end{figure}

The TES is located between $\pm$0.15 eV and its DOS are predicted to have a strong peak at the energy of $-$0.06~eV below the Fermi energy as shown in Fig.~2(b) and Fig.~5. The STS measurements on a few well-ordered edges consistently exhibit the clear enhancement in $dI/dV$ at this energy as also shown in Fig.~2(b) (the red line). We obtained STS ($dI/dV$) spectra sequentially along the line across one zigzag edge of a Bi(111) BL island [Fig.~6]. We notice apparent intensity modulations of the prominent spectral features at the edge region; the reduction $\beta$ and the appearance of a new peak at about $-$0.05~eV [see also Figs.~2(b) and 6]. The $\alpha$ state exhibits a similar reduction near the edge but also an enhancement way from the edge. The intensity modulations of $\alpha$ and other minor higher energy features (all valence band contribution of Bi(111) BL) follow the standing wave behavior due to the electron backscattering by the edge, which will be described elsewhere in detail. While such signal modulations appear for a wide distance of 10 nm or so from the edges, the newly observed state at $-$0.05~eV appear only at the edge region and its energy is fully consistent with the calculation for the TES [Figs.~2(b) and 2(d)].

For some edges probed, the edge enhancement of $\gamma$ becomes little bit more noticeable, which is consistent with what observed previously.\cite{yang_spatial_2012} However, we can clearly state that the distinct spectral feature at $-$0.05 eV is the unambiguous signature of the TES in good agreement with first-principles calculations. One can also note that the filling of the hybridization gap at the edge is due to the edge state DOS crossing this gap as shown in the calculation.

% Figure6
\begin{figure}[!pb]
\includegraphics{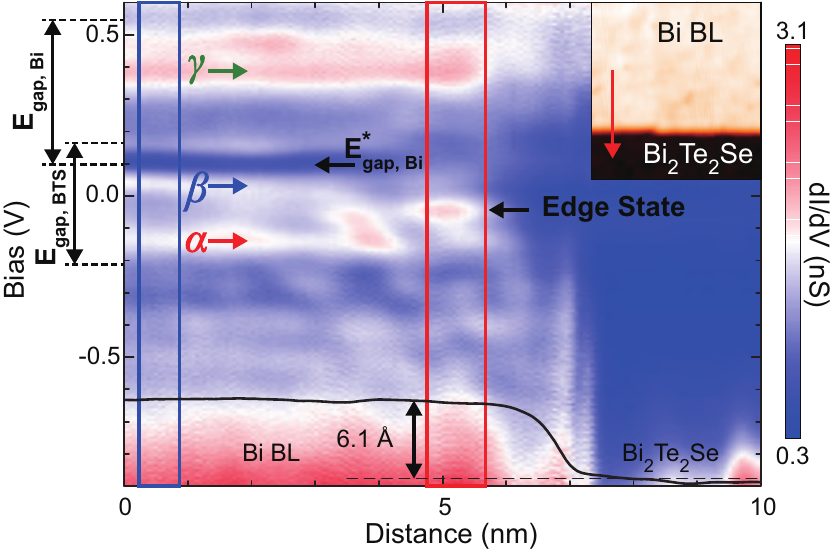}
\caption{The 2D plot of the STS ($dI/dV$) line scan along the arrow indicated in the inset, which crosses one zigzag edge of a Bi(111) BL island. The topography image in the inset covers an area of 20~nm~$\times$~20~nm. Averaged $dI/dV$ curves in the blue and red rectangular regions, representing the center and edge parts of the island, are shown in Fig.~2(b). Three major spectral features of the Bi(111) bilayer film and the distinct edge state are indicated by arrows. The topographic profile obtained during the STS measurement is included at the bottom of the figure.}
\end{figure}

The edge state of the DOS is further corroborated by the spatial STS ($dI/dV$) maps for a Bi(111) island. Such STS maps in Fig.~7 clearly indicate the enhanced contrast along one edge of a Bi(111) BL island within about 2--3~nm from the edge. The energy range for this enhancement is well confined between $-$0.1~V and Fermi level [Figs.~2(b) and 7(d)], which agrees well with that of the $dI/dV$ spectra and that of our calculation for the edge state. We thus conclude that the edge state is localized within this energy range and within about 2--3 nm from the step edge [see Figs.~6 and 7]. The reason for the edge state appearance with double lines in this map is not completely clear but is thought to be related to the electron standing wave of the valence band of the Bi film overlapped in this energy range.

% Figure7
\begin{figure}[!pt]
\includegraphics{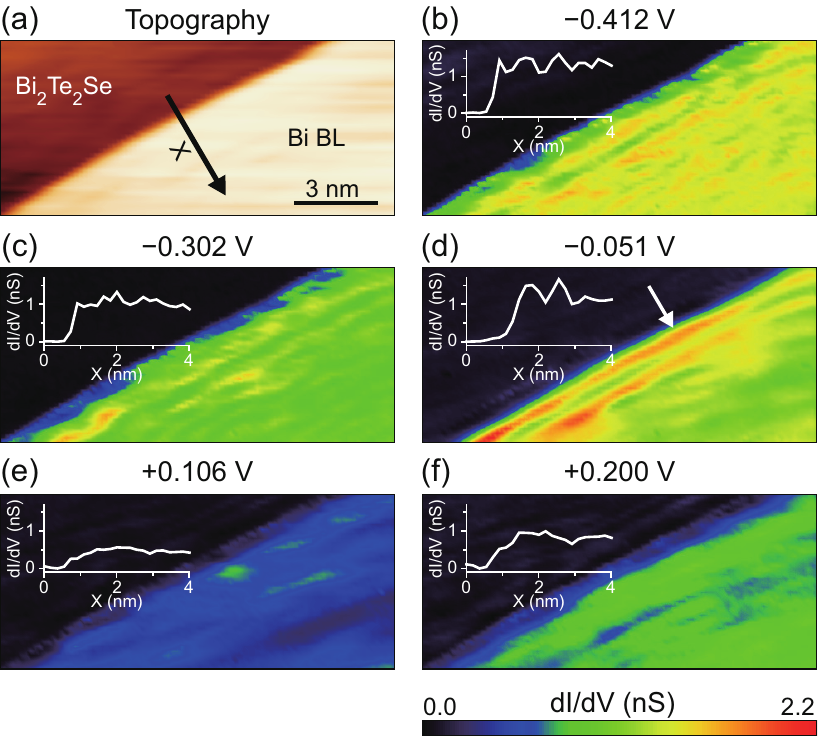}
\caption{(a) STM topographic image (14.1~nm~$\times$~6.2~nm) of a Bi(111) island on Bi$_{2}$Te$_{2}$Se. STS ($dI/dV$) maps at (b) $-$412, (c) $-$0.302, (d) $-$0.051, (e) $+$0.106 and (f) $+$0.200~V for the area of (a). The profile of the STS maps along the arrow (X) in (a) shows as inset each figure. Around $-0.05$~V, the Bi(111) BL island edges show a pronounced contrast enhancement in the STS maps indicating the edge state DOS.}
\end{figure}

In sharp contrast to the present result, the previous STS work showed the marginally enhanced $dI/dV$ contrast at a very different energy range in the empty states.\cite{yang_spatial_2012} This energy corresponds to $\gamma$ near the CBM of a floating Bi(111) BL and also to the highest energy for the edge state predicted for a floating BL. However, even for the floating BL, the edge state DOS at this energy is very low. Moreover, for the epitaxial BL, this energy is well out of the band gap due to the closing of the original gap and the formation of the hybridization gap. That is, if this enhancement is due to the TES, we have a fully unoccupied edge state well out of the band gap, which is apparently out of sense. The previous result itself could be reproduced in the present measurement as the intensity modulation of $\gamma$, the interfacial state. It is naturally expected that an interfacial state could get a strong perturbation near an edge of the island.
\\

\section{Conclusions}
We investigated the electronic states at the interfaces and along the zigzag step edge of a single Bi(111) bilayer film grown on Bi$_{2}$Te$_{2}$Se by STM, STS measurements and first-principle calculations. The strong interfacial interaction with the substrate induces a large splitting of the degenerate valence bands of Bi(111) and the band hybridization with those of the substrate. The hybridization leads to the formation of a characteristic interfacial state and a small hybridization energy gap above the Fermi level for the Bi film. Irrespective of such a strong interaction with the substrate, the band inversion of the Bi film and the edge state are shown to be robust. The edge state along the zigzag edge exhibits a substantial energy shift. This edge state signature was unambiguously resolved in the STS spectra and spatial maps. The interfacial state is also affected near the edges, which were misinterpreted as the sign of the edge state in the previous work.\cite{yang_spatial_2012} While our work may represents the clear real space observation of the edge electronic state of a 2D TI, which corresponds to a QSH state, the topological or helical nature of the edge state has to be confirmed further, for example, by a spin-polarized tunneling experiment.\cite{Spin_das_2011} Microscopic understanding of the nature of the edge states will help explore exotic low dimensional physics such as the helical Luttinger liquid and emerging physics from the coexistence of the 2D and 3D TI with their 1D and 2D TES's in proximity. 

\section{Acknowledgements}
This work was supported by Institute for Basic Science (IBS) through the Center for Artificial Low Dimensional Electronic Systems, the SRC Center for Topological Matter (2011-0030789) and the Basic Science Research program (2012-013838). We appreciate the help of H.~Ishikawa and K.~Sakamoto in performing the ARPES measurement.


\begin{thebibliography}{30}%
\makeatletter
\providecommand \@ifxundefined [1]{%
 \@ifx{#1\undefined}
}%
\providecommand \@ifnum [1]{%
 \ifnum #1\expandafter \@firstoftwo
 \else \expandafter \@secondoftwo
 \fi
}%
\providecommand \@ifx [1]{%
 \ifx #1\expandafter \@firstoftwo
 \else \expandafter \@secondoftwo
 \fi
}%
\providecommand \natexlab [1]{#1}%
\providecommand \enquote  [1]{``#1''}%
\providecommand \bibnamefont  [1]{#1}%
\providecommand \bibfnamefont [1]{#1}%
\providecommand \citenamefont [1]{#1}%
\providecommand \href@noop [0]{\@secondoftwo}%
\providecommand \href [0]{\begingroup \@sanitize@url \@href}%
\providecommand \@href[1]{\@@startlink{#1}\@@href}%
\providecommand \@@href[1]{\endgroup#1\@@endlink}%
\providecommand \@sanitize@url [0]{\catcode `\\12\catcode `\$12\catcode
  `\&12\catcode `\#12\catcode `\^12\catcode `\_12\catcode `\%12\relax}%
\providecommand \@@startlink[1]{}%
\providecommand \@@endlink[0]{}%
\providecommand \url  [0]{\begingroup\@sanitize@url \@url }%
\providecommand \@url [1]{\endgroup\@href {#1}{\urlprefix }}%
\providecommand \urlprefix  [0]{URL }%
\providecommand \Eprint [0]{\href }%
\providecommand \doibase [0]{http://dx.doi.org/}%
\providecommand \selectlanguage [0]{\@gobble}%
\providecommand \bibinfo  [0]{\@secondoftwo}%
\providecommand \bibfield  [0]{\@secondoftwo}%
\providecommand \translation [1]{[#1]}%
\providecommand \BibitemOpen [0]{}%
\providecommand \bibitemStop [0]{}%
\providecommand \bibitemNoStop [0]{.\EOS\space}%
\providecommand \EOS [0]{\spacefactor3000\relax}%
\providecommand \BibitemShut  [1]{\csname bibitem#1\endcsname}%
\let\auto@bib@innerbib\@empty
%</preamble>
\bibitem [{\citenamefont {Bernevig}\ \emph {et~al.}(2006)\citenamefont
  {Bernevig}, \citenamefont {Hughes},\ and\ \citenamefont
  {Zhang}}]{bernevig_quantum_2006}%
  \BibitemOpen
  \bibfield  {author} {\bibinfo {author} {\bibfnamefont {B.~A.}\ \bibnamefont
  {Bernevig}}, \bibinfo {author} {\bibfnamefont {T.~L.}\ \bibnamefont
  {Hughes}}, \ and\ \bibinfo {author} {\bibfnamefont {S.-C.}\ \bibnamefont
  {Zhang}},\ }\href {\doibase 10.1126/science.1133734} {\bibfield  {journal}
  {\bibinfo  {journal} {Science}\ }\textbf {\bibinfo {volume} {314}},\ \bibinfo
  {pages} {1757} (\bibinfo {year} {2006})}\BibitemShut {NoStop}%
\bibitem [{\citenamefont {K\"{o}nig}\ \emph {et~al.}(2007)\citenamefont
  {K\"{o}nig}, \citenamefont {Wiedmann}, \citenamefont {Brüne}, \citenamefont
  {Roth}, \citenamefont {Buhmann}, \citenamefont {Molenkamp}, \citenamefont
  {Qi},\ and\ \citenamefont {Zhang}}]{konig_quantum_2007}%
  \BibitemOpen
  \bibfield  {author} {\bibinfo {author} {\bibfnamefont {M.}~\bibnamefont
  {K\"{o}nig}}, \bibinfo {author} {\bibfnamefont {S.}~\bibnamefont {Wiedmann}},
  \bibinfo {author} {\bibfnamefont {C.}~\bibnamefont {Brüne}}, \bibinfo
  {author} {\bibfnamefont {A.}~\bibnamefont {Roth}}, \bibinfo {author}
  {\bibfnamefont {H.}~\bibnamefont {Buhmann}}, \bibinfo {author} {\bibfnamefont
  {L.~W.}\ \bibnamefont {Molenkamp}}, \bibinfo {author} {\bibfnamefont {X.-L.}\
  \bibnamefont {Qi}}, \ and\ \bibinfo {author} {\bibfnamefont {S.-C.}\
  \bibnamefont {Zhang}},\ }\href {\doibase 10.1126/science.1148047} {\bibfield
  {journal} {\bibinfo  {journal} {Science}\ }\textbf {\bibinfo {volume}
  {318}},\ \bibinfo {pages} {766} (\bibinfo {year} {2007})}\BibitemShut
  {NoStop}%
\bibitem [{\citenamefont {Hsieh}\ \emph {et~al.}(2008)\citenamefont {Hsieh},
  \citenamefont {Qian}, \citenamefont {Wray}, \citenamefont {Xia},
  \citenamefont {Hor}, \citenamefont {Cava},\ and\ \citenamefont
  {Hasan}}]{hsieh_topological_2008}%
  \BibitemOpen
  \bibfield  {author} {\bibinfo {author} {\bibfnamefont {D.}~\bibnamefont
  {Hsieh}}, \bibinfo {author} {\bibfnamefont {D.}~\bibnamefont {Qian}},
  \bibinfo {author} {\bibfnamefont {L.}~\bibnamefont {Wray}}, \bibinfo {author}
  {\bibfnamefont {Y.}~\bibnamefont {Xia}}, \bibinfo {author} {\bibfnamefont
  {Y.~S.}\ \bibnamefont {Hor}}, \bibinfo {author} {\bibfnamefont {R.~J.}\
  \bibnamefont {Cava}}, \ and\ \bibinfo {author} {\bibfnamefont {M.~Z.}\
  \bibnamefont {Hasan}},\ }\href {\doibase 10.1038/nature06843} {\bibfield
  {journal} {\bibinfo  {journal} {Nature (London)}\ }\textbf {\bibinfo {volume}
  {452}},\ \bibinfo {pages} {970} (\bibinfo {year} {2008})}\BibitemShut
  {NoStop}%
\bibitem [{\citenamefont {Chen}\ \emph {et~al.}(2009)\citenamefont {Chen},
  \citenamefont {Analytis}, \citenamefont {Chu}, \citenamefont {Liu},
  \citenamefont {Mo}, \citenamefont {Qi}, \citenamefont {Zhang}, \citenamefont
  {Lu}, \citenamefont {Dai}, \citenamefont {Fang}, \citenamefont {Zhang},
  \citenamefont {Fisher}, \citenamefont {Hussain},\ and\ \citenamefont
  {Shen}}]{chen_experimental_2009}%
  \BibitemOpen
  \bibfield  {author} {\bibinfo {author} {\bibfnamefont {Y.~L.}\ \bibnamefont
  {Chen}}, \bibinfo {author} {\bibfnamefont {J.~G.}\ \bibnamefont {Analytis}},
  \bibinfo {author} {\bibfnamefont {J.-H.}\ \bibnamefont {Chu}}, \bibinfo
  {author} {\bibfnamefont {Z.~K.}\ \bibnamefont {Liu}}, \bibinfo {author}
  {\bibfnamefont {S.-K.}\ \bibnamefont {Mo}}, \bibinfo {author} {\bibfnamefont
  {X.~L.}\ \bibnamefont {Qi}}, \bibinfo {author} {\bibfnamefont {H.~J.}\
  \bibnamefont {Zhang}}, \bibinfo {author} {\bibfnamefont {D.~H.}\ \bibnamefont
  {Lu}}, \bibinfo {author} {\bibfnamefont {X.}~\bibnamefont {Dai}}, \bibinfo
  {author} {\bibfnamefont {Z.}~\bibnamefont {Fang}}, \bibinfo {author}
  {\bibfnamefont {S.~C.}\ \bibnamefont {Zhang}}, \bibinfo {author}
  {\bibfnamefont {I.~R.}\ \bibnamefont {Fisher}}, \bibinfo {author}
  {\bibfnamefont {Z.}~\bibnamefont {Hussain}}, \ and\ \bibinfo {author}
  {\bibfnamefont {Z.-X.}\ \bibnamefont {Shen}},\ }\href {\doibase
  10.1126/science.1173034} {\bibfield  {journal} {\bibinfo  {journal}
  {Science}\ }\textbf {\bibinfo {volume} {325}},\ \bibinfo {pages} {178}
  (\bibinfo {year} {2009})}\BibitemShut {NoStop}%
\bibitem [{\citenamefont {Hasan}\ and\ \citenamefont
  {Kane}(2010)}]{hasan_colloquium:_2010}%
  \BibitemOpen
  \bibfield  {author} {\bibinfo {author} {\bibfnamefont {M.~Z.}\ \bibnamefont
  {Hasan}}\ and\ \bibinfo {author} {\bibfnamefont {C.~L.}\ \bibnamefont
  {Kane}},\ }\href {\doibase 10.1103/RevModPhys.82.3045} {\bibfield  {journal}
  {\bibinfo  {journal} {Rev. Mod. Phys.}\ }\textbf {\bibinfo {volume} {82}},\
  \bibinfo {pages} {3045} (\bibinfo {year} {2010})}\BibitemShut {NoStop}%
\bibitem [{\citenamefont {Kuroda}\ \emph {et~al.}(2010)\citenamefont {Kuroda},
  \citenamefont {Ye}, \citenamefont {Kimura}, \citenamefont {Eremeev},
  \citenamefont {Krasovskii}, \citenamefont {Chulkov}, \citenamefont {Ueda},
  \citenamefont {Miyamoto}, \citenamefont {Okuda}, \citenamefont {Shimada},
  \citenamefont {Namatame},\ and\ \citenamefont
  {Taniguchi}}]{kuroda_experimental_2010}%
  \BibitemOpen
  \bibfield  {author} {\bibinfo {author} {\bibfnamefont {K.}~\bibnamefont
  {Kuroda}}, \bibinfo {author} {\bibfnamefont {M.}~\bibnamefont {Ye}}, \bibinfo
  {author} {\bibfnamefont {A.}~\bibnamefont {Kimura}}, \bibinfo {author}
  {\bibfnamefont {S.~V.}\ \bibnamefont {Eremeev}}, \bibinfo {author}
  {\bibfnamefont {E.~E.}\ \bibnamefont {Krasovskii}}, \bibinfo {author}
  {\bibfnamefont {E.~V.}\ \bibnamefont {Chulkov}}, \bibinfo {author}
  {\bibfnamefont {Y.}~\bibnamefont {Ueda}}, \bibinfo {author} {\bibfnamefont
  {K.}~\bibnamefont {Miyamoto}}, \bibinfo {author} {\bibfnamefont
  {T.}~\bibnamefont {Okuda}}, \bibinfo {author} {\bibfnamefont
  {K.}~\bibnamefont {Shimada}}, \bibinfo {author} {\bibfnamefont
  {H.}~\bibnamefont {Namatame}}, \ and\ \bibinfo {author} {\bibfnamefont
  {M.}~\bibnamefont {Taniguchi}},\ }\href {\doibase
  10.1103/PhysRevLett.105.146801} {\bibfield  {journal} {\bibinfo  {journal}
  {Phys. Rev. Lett.}\ }\textbf {\bibinfo {volume} {105}},\ \bibinfo {pages}
  {146801} (\bibinfo {year} {2010})}\BibitemShut {NoStop}%
\bibitem [{\citenamefont {Souma}\ \emph {et~al.}(2012)\citenamefont {Souma},
  \citenamefont {Eto}, \citenamefont {Nomura}, \citenamefont {Nakayama},
  \citenamefont {Sato}, \citenamefont {Takahashi}, \citenamefont {Segawa},\
  and\ \citenamefont {Ando}}]{souma_topological_2012}%
  \BibitemOpen
  \bibfield  {author} {\bibinfo {author} {\bibfnamefont {S.}~\bibnamefont
  {Souma}}, \bibinfo {author} {\bibfnamefont {K.}~\bibnamefont {Eto}}, \bibinfo
  {author} {\bibfnamefont {M.}~\bibnamefont {Nomura}}, \bibinfo {author}
  {\bibfnamefont {K.}~\bibnamefont {Nakayama}}, \bibinfo {author}
  {\bibfnamefont {T.}~\bibnamefont {Sato}}, \bibinfo {author} {\bibfnamefont
  {T.}~\bibnamefont {Takahashi}}, \bibinfo {author} {\bibfnamefont
  {K.}~\bibnamefont {Segawa}}, \ and\ \bibinfo {author} {\bibfnamefont
  {Y.}~\bibnamefont {Ando}},\ }\href {\doibase 10.1103/PhysRevLett.108.116801}
  {\bibfield  {journal} {\bibinfo  {journal} {Phys. Rev. Lett.}\ }\textbf
  {\bibinfo {volume} {108}},\ \bibinfo {pages} {116801} (\bibinfo {year}
  {2012})}\BibitemShut {NoStop}%
\bibitem [{\citenamefont {Al-Sawai}\ \emph {et~al.}(2010)\citenamefont
  {Al-Sawai}, \citenamefont {Lin}, \citenamefont {Markiewicz}, \citenamefont
  {Wray}, \citenamefont {Xia}, \citenamefont {Xu}, \citenamefont {Hasan},\ and\
  \citenamefont {Bansil}}]{al-sawai_topological_2010}%
  \BibitemOpen
  \bibfield  {author} {\bibinfo {author} {\bibfnamefont {W.}~\bibnamefont
  {Al-Sawai}}, \bibinfo {author} {\bibfnamefont {H.}~\bibnamefont {Lin}},
  \bibinfo {author} {\bibfnamefont {R.~S.}\ \bibnamefont {Markiewicz}},
  \bibinfo {author} {\bibfnamefont {L.~A.}\ \bibnamefont {Wray}}, \bibinfo
  {author} {\bibfnamefont {Y.}~\bibnamefont {Xia}}, \bibinfo {author}
  {\bibfnamefont {S.-Y.}\ \bibnamefont {Xu}}, \bibinfo {author} {\bibfnamefont
  {M.~Z.}\ \bibnamefont {Hasan}}, \ and\ \bibinfo {author} {\bibfnamefont
  {A.}~\bibnamefont {Bansil}},\ }\href {\doibase 10.1103/PhysRevB.82.125208}
  {\bibfield  {journal} {\bibinfo  {journal} {Phys. Rev. B}\ }\textbf {\bibinfo
  {volume} {82}},\ \bibinfo {pages} {125208} (\bibinfo {year}
  {2010})}\BibitemShut {NoStop}%
\bibitem [{\citenamefont {Liu}\ \emph {et~al.}(2011{\natexlab{a}})\citenamefont
  {Liu}, \citenamefont {Lee}, \citenamefont {Kondo}, \citenamefont {Mun},
  \citenamefont {Caudle}, \citenamefont {Harmon}, \citenamefont {Bud'ko},
  \citenamefont {Canfield},\ and\ \citenamefont
  {Kaminski}}]{liu_metallic_2011}%
  \BibitemOpen
  \bibfield  {author} {\bibinfo {author} {\bibfnamefont {C.}~\bibnamefont
  {Liu}}, \bibinfo {author} {\bibfnamefont {Y.}~\bibnamefont {Lee}}, \bibinfo
  {author} {\bibfnamefont {T.}~\bibnamefont {Kondo}}, \bibinfo {author}
  {\bibfnamefont {E.~D.}\ \bibnamefont {Mun}}, \bibinfo {author} {\bibfnamefont
  {M.}~\bibnamefont {Caudle}}, \bibinfo {author} {\bibfnamefont {B.~N.}\
  \bibnamefont {Harmon}}, \bibinfo {author} {\bibfnamefont {S.~L.}\
  \bibnamefont {Bud'ko}}, \bibinfo {author} {\bibfnamefont {P.~C.}\
  \bibnamefont {Canfield}}, \ and\ \bibinfo {author} {\bibfnamefont
  {A.}~\bibnamefont {Kaminski}},\ }\href {\doibase 10.1103/PhysRevB.83.205133}
  {\bibfield  {journal} {\bibinfo  {journal} {Phys. Rev. B}\ }\textbf {\bibinfo
  {volume} {83}},\ \bibinfo {pages} {205133} (\bibinfo {year}
  {2011}{\natexlab{a}})}\BibitemShut {NoStop}%
\bibitem [{\citenamefont {Sun}\ \emph {et~al.}(2010)\citenamefont {Sun},
  \citenamefont {Chen}, \citenamefont {Yunoki}, \citenamefont {Li},\ and\
  \citenamefont {Li}}]{sun_new_2010}%
  \BibitemOpen
  \bibfield  {author} {\bibinfo {author} {\bibfnamefont {Y.}~\bibnamefont
  {Sun}}, \bibinfo {author} {\bibfnamefont {X.-Q.}\ \bibnamefont {Chen}},
  \bibinfo {author} {\bibfnamefont {S.}~\bibnamefont {Yunoki}}, \bibinfo
  {author} {\bibfnamefont {D.}~\bibnamefont {Li}}, \ and\ \bibinfo {author}
  {\bibfnamefont {Y.}~\bibnamefont {Li}},\ }\href {\doibase
  10.1103/PhysRevLett.105.216406} {\bibfield  {journal} {\bibinfo  {journal}
  {Phys. Rev. Lett.}\ }\textbf {\bibinfo {volume} {105}},\ \bibinfo {pages}
  {216406} (\bibinfo {year} {2010})}\BibitemShut {NoStop}%
\bibitem [{\citenamefont {Zhang}\ \emph {et~al.}(2009)\citenamefont {Zhang},
  \citenamefont {Liu}, \citenamefont {Qi}, \citenamefont {Dai}, \citenamefont
  {Fang},\ and\ \citenamefont {Zhang}}]{zhang_topological_2009}%
  \BibitemOpen
  \bibfield  {author} {\bibinfo {author} {\bibfnamefont {H.}~\bibnamefont
  {Zhang}}, \bibinfo {author} {\bibfnamefont {C.-X.}\ \bibnamefont {Liu}},
  \bibinfo {author} {\bibfnamefont {X.-L.}\ \bibnamefont {Qi}}, \bibinfo
  {author} {\bibfnamefont {X.}~\bibnamefont {Dai}}, \bibinfo {author}
  {\bibfnamefont {Z.}~\bibnamefont {Fang}}, \ and\ \bibinfo {author}
  {\bibfnamefont {S.-C.}\ \bibnamefont {Zhang}},\ }\href {\doibase
  10.1038/nphys1270} {\bibfield  {journal} {\bibinfo  {journal} {Nature Phys.}\
  }\textbf {\bibinfo {volume} {5}},\ \bibinfo {pages} {438} (\bibinfo {year}
  {2009})}\BibitemShut {NoStop}%
\bibitem [{\citenamefont {Murakami}\ \emph {et~al.}(2003)\citenamefont
  {Murakami}, \citenamefont {Nagaosa},\ and\ \citenamefont
  {Zhang}}]{murakami_dissipationless_2003}%
  \BibitemOpen
  \bibfield  {author} {\bibinfo {author} {\bibfnamefont {S.}~\bibnamefont
  {Murakami}}, \bibinfo {author} {\bibfnamefont {N.}~\bibnamefont {Nagaosa}}, \
  and\ \bibinfo {author} {\bibfnamefont {S.-C.}\ \bibnamefont {Zhang}},\ }\href
  {\doibase 10.1126/science.1087128} {\bibfield  {journal} {\bibinfo  {journal}
  {Science}\ }\textbf {\bibinfo {volume} {301}},\ \bibinfo {pages} {1348}
  (\bibinfo {year} {2003})}\BibitemShut {NoStop}%
\bibitem [{\citenamefont {Sinova}\ \emph {et~al.}(2004)\citenamefont {Sinova},
  \citenamefont {Culcer}, \citenamefont {Niu}, \citenamefont {Sinitsyn},
  \citenamefont {Jungwirth},\ and\ \citenamefont
  {MacDonald}}]{sinova_universal_2004}%
  \BibitemOpen
  \bibfield  {author} {\bibinfo {author} {\bibfnamefont {J.}~\bibnamefont
  {Sinova}}, \bibinfo {author} {\bibfnamefont {D.}~\bibnamefont {Culcer}},
  \bibinfo {author} {\bibfnamefont {Q.}~\bibnamefont {Niu}}, \bibinfo {author}
  {\bibfnamefont {N.~A.}\ \bibnamefont {Sinitsyn}}, \bibinfo {author}
  {\bibfnamefont {T.}~\bibnamefont {Jungwirth}}, \ and\ \bibinfo {author}
  {\bibfnamefont {A.~H.}\ \bibnamefont {MacDonald}},\ }\href {\doibase
  10.1103/PhysRevLett.92.126603} {\bibfield  {journal} {\bibinfo  {journal}
  {Phys. Rev. Lett.}\ }\textbf {\bibinfo {volume} {92}},\ \bibinfo {pages}
  {126603} (\bibinfo {year} {2004})}\BibitemShut {NoStop}%
\bibitem [{\citenamefont {Liu}\ \emph {et~al.}(2008)\citenamefont {Liu},
  \citenamefont {Hughes}, \citenamefont {Qi}, \citenamefont {Wang},\ and\
  \citenamefont {Zhang}}]{InAs_theory}%
  \BibitemOpen
  \bibfield  {author} {\bibinfo {author} {\bibfnamefont {C.}~\bibnamefont
  {Liu}}, \bibinfo {author} {\bibfnamefont {T.~L.}\ \bibnamefont {Hughes}},
  \bibinfo {author} {\bibfnamefont {X.-L.}\ \bibnamefont {Qi}}, \bibinfo
  {author} {\bibfnamefont {K.}~\bibnamefont {Wang}}, \ and\ \bibinfo {author}
  {\bibfnamefont {S.-C.}\ \bibnamefont {Zhang}},\ }\href {\doibase
  10.1103/PhysRevLett.100.236601} {\bibfield  {journal} {\bibinfo  {journal}
  {Phys. Rev. Lett.}\ }\textbf {\bibinfo {volume} {100}},\ \bibinfo {pages}
  {236601} (\bibinfo {year} {2008})}\BibitemShut {NoStop}%
\bibitem [{\citenamefont {Knez}\ \emph {et~al.}(2011)\citenamefont {Knez},
  \citenamefont {Du},\ and\ \citenamefont {Sullivan}}]{InAs_expt}%
  \BibitemOpen
  \bibfield  {author} {\bibinfo {author} {\bibfnamefont {I.}~\bibnamefont
  {Knez}}, \bibinfo {author} {\bibfnamefont {R.-R.}\ \bibnamefont {Du}}, \ and\
  \bibinfo {author} {\bibfnamefont {G.}~\bibnamefont {Sullivan}},\ }\href
  {\doibase 10.1103/PhysRevLett.107.136603} {\bibfield  {journal} {\bibinfo
  {journal} {Phys. Rev. Lett.}\ }\textbf {\bibinfo {volume} {107}},\ \bibinfo
  {pages} {136603} (\bibinfo {year} {2011})}\BibitemShut {NoStop}%
\bibitem [{\citenamefont {Murakami}(2006)}]{murakami_quantum_2006}%
  \BibitemOpen
  \bibfield  {author} {\bibinfo {author} {\bibfnamefont {S.}~\bibnamefont
  {Murakami}},\ }\href {\doibase 10.1103/PhysRevLett.97.236805} {\bibfield
  {journal} {\bibinfo  {journal} {Phys. Rev. Lett.}\ }\textbf {\bibinfo
  {volume} {97}},\ \bibinfo {pages} {236805} (\bibinfo {year}
  {2006})}\BibitemShut {NoStop}%
\bibitem [{\citenamefont {Wada}\ \emph {et~al.}(2011)\citenamefont {Wada},
  \citenamefont {Murakami}, \citenamefont {Freimuth},\ and\ \citenamefont
  {Bihlmayer}}]{wada_localized_2011}%
  \BibitemOpen
  \bibfield  {author} {\bibinfo {author} {\bibfnamefont {M.}~\bibnamefont
  {Wada}}, \bibinfo {author} {\bibfnamefont {S.}~\bibnamefont {Murakami}},
  \bibinfo {author} {\bibfnamefont {F.}~\bibnamefont {Freimuth}}, \ and\
  \bibinfo {author} {\bibfnamefont {G.}~\bibnamefont {Bihlmayer}},\ }\href
  {\doibase 10.1103/PhysRevB.83.121310} {\bibfield  {journal} {\bibinfo
  {journal} {Phys. Rev. B}\ }\textbf {\bibinfo {volume} {83}},\ \bibinfo
  {pages} {121310} (\bibinfo {year} {2011})}\BibitemShut {NoStop}%
\bibitem [{\citenamefont {Sakamoto}\ \emph {et~al.}(2010)\citenamefont
  {Sakamoto}, \citenamefont {Hirahara}, \citenamefont {Miyazaki}, \citenamefont
  {Kimura},\ and\ \citenamefont {Hasegawa}}]{sakamoto_spectroscopic_2010}%
  \BibitemOpen
  \bibfield  {author} {\bibinfo {author} {\bibfnamefont {Y.}~\bibnamefont
  {Sakamoto}}, \bibinfo {author} {\bibfnamefont {T.}~\bibnamefont {Hirahara}},
  \bibinfo {author} {\bibfnamefont {H.}~\bibnamefont {Miyazaki}}, \bibinfo
  {author} {\bibfnamefont {S.-I.}\ \bibnamefont {Kimura}}, \ and\ \bibinfo
  {author} {\bibfnamefont {S.}~\bibnamefont {Hasegawa}},\ }\href {\doibase
  10.1103/PhysRevB.81.165432} {\bibfield  {journal} {\bibinfo  {journal} {Phys.
  Rev. B}\ }\textbf {\bibinfo {volume} {81}},\ \bibinfo {pages} {165432}
  (\bibinfo {year} {2010})}\BibitemShut {NoStop}%
\bibitem [{\citenamefont {Hirahara}\ \emph {et~al.}(2011)\citenamefont
  {Hirahara}, \citenamefont {Bihlmayer}, \citenamefont {Sakamoto},
  \citenamefont {Yamada}, \citenamefont {Miyazaki}, \citenamefont {Kimura},
  \citenamefont {Bl\"ugel},\ and\ \citenamefont
  {Hasegawa}}]{hirahara_interfacing_2011}%
  \BibitemOpen
  \bibfield  {author} {\bibinfo {author} {\bibfnamefont {T.}~\bibnamefont
  {Hirahara}}, \bibinfo {author} {\bibfnamefont {G.}~\bibnamefont {Bihlmayer}},
  \bibinfo {author} {\bibfnamefont {Y.}~\bibnamefont {Sakamoto}}, \bibinfo
  {author} {\bibfnamefont {M.}~\bibnamefont {Yamada}}, \bibinfo {author}
  {\bibfnamefont {H.}~\bibnamefont {Miyazaki}}, \bibinfo {author}
  {\bibfnamefont {S.-i.}\ \bibnamefont {Kimura}}, \bibinfo {author}
  {\bibfnamefont {S.}~\bibnamefont {Bl\"ugel}}, \ and\ \bibinfo {author}
  {\bibfnamefont {S.}~\bibnamefont {Hasegawa}},\ }\href {\doibase
  10.1103/PhysRevLett.107.166801} {\bibfield  {journal} {\bibinfo  {journal}
  {Phys. Rev. Lett.}\ }\textbf {\bibinfo {volume} {107}},\ \bibinfo {pages}
  {166801} (\bibinfo {year} {2011})}\BibitemShut {NoStop}%
\bibitem [{\citenamefont {Liu}\ \emph {et~al.}(2011{\natexlab{b}})\citenamefont
  {Liu}, \citenamefont {Liu}, \citenamefont {Wu}, \citenamefont {Duan},
  \citenamefont {Liu},\ and\ \citenamefont {Wu}}]{liu_stable_2011}%
  \BibitemOpen
  \bibfield  {author} {\bibinfo {author} {\bibfnamefont {Z.}~\bibnamefont
  {Liu}}, \bibinfo {author} {\bibfnamefont {C.-X.}\ \bibnamefont {Liu}},
  \bibinfo {author} {\bibfnamefont {Y.-S.}\ \bibnamefont {Wu}}, \bibinfo
  {author} {\bibfnamefont {W.-H.}\ \bibnamefont {Duan}}, \bibinfo {author}
  {\bibfnamefont {F.}~\bibnamefont {Liu}}, \ and\ \bibinfo {author}
  {\bibfnamefont {J.}~\bibnamefont {Wu}},\ }\href {\doibase
  10.1103/PhysRevLett.107.136805} {\bibfield  {journal} {\bibinfo  {journal}
  {Phys. Rev. Lett.}\ }\textbf {\bibinfo {volume} {107}},\ \bibinfo {pages}
  {136805} (\bibinfo {year} {2011}{\natexlab{b}})}\BibitemShut {NoStop}%
\bibitem [{\citenamefont {Jnawali}\ \emph {et~al.}(2012)\citenamefont
  {Jnawali}, \citenamefont {Klein}, \citenamefont {Wagner}, \citenamefont
  {Hattab}, \citenamefont {Zahl}, \citenamefont {Acharya}, \citenamefont
  {Sutter}, \citenamefont {Lorke},\ and\ \citenamefont {Horn-von
  Hoegen}}]{jnawali_manipulation_2012}%
  \BibitemOpen
  \bibfield  {author} {\bibinfo {author} {\bibfnamefont {G.}~\bibnamefont
  {Jnawali}}, \bibinfo {author} {\bibfnamefont {C.}~\bibnamefont {Klein}},
  \bibinfo {author} {\bibfnamefont {T.}~\bibnamefont {Wagner}}, \bibinfo
  {author} {\bibfnamefont {H.}~\bibnamefont {Hattab}}, \bibinfo {author}
  {\bibfnamefont {P.}~\bibnamefont {Zahl}}, \bibinfo {author} {\bibfnamefont
  {D.~P.}\ \bibnamefont {Acharya}}, \bibinfo {author} {\bibfnamefont
  {P.}~\bibnamefont {Sutter}}, \bibinfo {author} {\bibfnamefont
  {A.}~\bibnamefont {Lorke}}, \ and\ \bibinfo {author} {\bibfnamefont
  {M.}~\bibnamefont {Horn-von Hoegen}},\ }\href {\doibase
  10.1103/PhysRevLett.108.266804} {\bibfield  {journal} {\bibinfo  {journal}
  {Phys. Rev. Lett.}\ }\textbf {\bibinfo {volume} {108}},\ \bibinfo {pages}
  {266804} (\bibinfo {year} {2012})}\BibitemShut {NoStop}%
\bibitem [{\citenamefont {Yang}\ \emph {et~al.}(2012)\citenamefont {Yang},
  \citenamefont {Miao}, \citenamefont {Wang}, \citenamefont {Yao},
  \citenamefont {Zhu}, \citenamefont {Song}, \citenamefont {Wang},
  \citenamefont {Xu}, \citenamefont {Fedorov}, \citenamefont {Sun},
  \citenamefont {Zhang}, \citenamefont {Liu}, \citenamefont {Liu},
  \citenamefont {Qian}, \citenamefont {Gao},\ and\ \citenamefont
  {Jia}}]{yang_spatial_2012}%
  \BibitemOpen
  \bibfield  {author} {\bibinfo {author} {\bibfnamefont {F.}~\bibnamefont
  {Yang}}, \bibinfo {author} {\bibfnamefont {L.}~\bibnamefont {Miao}}, \bibinfo
  {author} {\bibfnamefont {Z.~F.}\ \bibnamefont {Wang}}, \bibinfo {author}
  {\bibfnamefont {M.-Y.}\ \bibnamefont {Yao}}, \bibinfo {author} {\bibfnamefont
  {F.}~\bibnamefont {Zhu}}, \bibinfo {author} {\bibfnamefont {Y.~R.}\
  \bibnamefont {Song}}, \bibinfo {author} {\bibfnamefont {M.-X.}\ \bibnamefont
  {Wang}}, \bibinfo {author} {\bibfnamefont {J.-P.}\ \bibnamefont {Xu}},
  \bibinfo {author} {\bibfnamefont {A.~V.}\ \bibnamefont {Fedorov}}, \bibinfo
  {author} {\bibfnamefont {Z.}~\bibnamefont {Sun}}, \bibinfo {author}
  {\bibfnamefont {G.~B.}\ \bibnamefont {Zhang}}, \bibinfo {author}
  {\bibfnamefont {C.}~\bibnamefont {Liu}}, \bibinfo {author} {\bibfnamefont
  {F.}~\bibnamefont {Liu}}, \bibinfo {author} {\bibfnamefont {D.}~\bibnamefont
  {Qian}}, \bibinfo {author} {\bibfnamefont {C.~L.}\ \bibnamefont {Gao}}, \
  and\ \bibinfo {author} {\bibfnamefont {J.-F.}\ \bibnamefont {Jia}},\ }\href
  {\doibase 10.1103/PhysRevLett.109.016801} {\bibfield  {journal} {\bibinfo
  {journal} {Phys. Rev. Lett.}\ }\textbf {\bibinfo {volume} {109}},\ \bibinfo
  {pages} {016801} (\bibinfo {year} {2012})}\BibitemShut {NoStop}%
\bibitem [{\citenamefont {Miao}\ \emph {et~al.}(2013)\citenamefont {Miao},
  \citenamefont {Wang}, \citenamefont {Ming}, \citenamefont {Yao},
  \citenamefont {Wang}, \citenamefont {Yang}, \citenamefont {Song},
  \citenamefont {Zhu}, \citenamefont {Fedorov}, \citenamefont {Sun},
  \citenamefont {Gao}, \citenamefont {Liu}, \citenamefont {Xue}, \citenamefont
  {Liu}, \citenamefont {Liu}, \citenamefont {Qian},\ and\ \citenamefont
  {Jia}}]{miao_2013}%
  \BibitemOpen
  \bibfield  {author} {\bibinfo {author} {\bibfnamefont {L.}~\bibnamefont
  {Miao}}, \bibinfo {author} {\bibfnamefont {Z.~F.}\ \bibnamefont {Wang}},
  \bibinfo {author} {\bibfnamefont {W.}~\bibnamefont {Ming}}, \bibinfo {author}
  {\bibfnamefont {M.-Y.}\ \bibnamefont {Yao}}, \bibinfo {author} {\bibfnamefont
  {M.}~\bibnamefont {Wang}}, \bibinfo {author} {\bibfnamefont {F.}~\bibnamefont
  {Yang}}, \bibinfo {author} {\bibfnamefont {Y.~R.}\ \bibnamefont {Song}},
  \bibinfo {author} {\bibfnamefont {F.}~\bibnamefont {Zhu}}, \bibinfo {author}
  {\bibfnamefont {A.~V.}\ \bibnamefont {Fedorov}}, \bibinfo {author}
  {\bibfnamefont {Z.}~\bibnamefont {Sun}}, \bibinfo {author} {\bibfnamefont
  {C.~L.}\ \bibnamefont {Gao}}, \bibinfo {author} {\bibfnamefont
  {C.}~\bibnamefont {Liu}}, \bibinfo {author} {\bibfnamefont {Q.-K.}\
  \bibnamefont {Xue}}, \bibinfo {author} {\bibfnamefont {C.-X.}\ \bibnamefont
  {Liu}}, \bibinfo {author} {\bibfnamefont {F.}~\bibnamefont {Liu}}, \bibinfo
  {author} {\bibfnamefont {D.}~\bibnamefont {Qian}}, \ and\ \bibinfo {author}
  {\bibfnamefont {J.-F.}\ \bibnamefont {Jia}},\ }\href {\doibase
  10.1073/pnas.1218104110} {\bibfield  {journal} {\bibinfo  {journal} {Proc.
  Natl. Acad. Sci. USA}\ }\textbf {\bibinfo {volume} {110}},\ \bibinfo {pages}
  {2758} (\bibinfo {year} {2013})}\BibitemShut {NoStop}%
\bibitem [{\citenamefont {Ren}\ \emph {et~al.}(2010)\citenamefont {Ren},
  \citenamefont {Taskin}, \citenamefont {Sasaki}, \citenamefont {Segawa},\ and\
  \citenamefont {Ando}}]{ren_large_2010}%
  \BibitemOpen
  \bibfield  {author} {\bibinfo {author} {\bibfnamefont {Z.}~\bibnamefont
  {Ren}}, \bibinfo {author} {\bibfnamefont {A.~A.}\ \bibnamefont {Taskin}},
  \bibinfo {author} {\bibfnamefont {S.}~\bibnamefont {Sasaki}}, \bibinfo
  {author} {\bibfnamefont {K.}~\bibnamefont {Segawa}}, \ and\ \bibinfo {author}
  {\bibfnamefont {Y.}~\bibnamefont {Ando}},\ }\href {\doibase
  10.1103/PhysRevB.82.241306} {\bibfield  {journal} {\bibinfo  {journal} {Phys.
  Rev. B}\ }\textbf {\bibinfo {volume} {82}},\ \bibinfo {pages} {241306}
  (\bibinfo {year} {2010})}\BibitemShut {NoStop}%
\bibitem [{\citenamefont {Alpichshev}\ \emph {et~al.}(2010)\citenamefont
  {Alpichshev}, \citenamefont {Analytis}, \citenamefont {Chu}, \citenamefont
  {Fisher}, \citenamefont {Chen}, \citenamefont {Shen}, \citenamefont {Fang},\
  and\ \citenamefont {Kapitulnik}}]{alpichshev_stm_2010}%
  \BibitemOpen
  \bibfield  {author} {\bibinfo {author} {\bibfnamefont {Z.}~\bibnamefont
  {Alpichshev}}, \bibinfo {author} {\bibfnamefont {J.~G.}\ \bibnamefont
  {Analytis}}, \bibinfo {author} {\bibfnamefont {J.-H.}\ \bibnamefont {Chu}},
  \bibinfo {author} {\bibfnamefont {I.~R.}\ \bibnamefont {Fisher}}, \bibinfo
  {author} {\bibfnamefont {Y.~L.}\ \bibnamefont {Chen}}, \bibinfo {author}
  {\bibfnamefont {Z.~X.}\ \bibnamefont {Shen}}, \bibinfo {author}
  {\bibfnamefont {A.}~\bibnamefont {Fang}}, \ and\ \bibinfo {author}
  {\bibfnamefont {A.}~\bibnamefont {Kapitulnik}},\ }\href {\doibase
  10.1103/PhysRevLett.104.016401} {\bibfield  {journal} {\bibinfo  {journal}
  {Phys. Rev. Lett.}\ }\textbf {\bibinfo {volume} {104}},\ \bibinfo {pages}
  {016401} (\bibinfo {year} {2010})}\BibitemShut {NoStop}%
\bibitem [{\citenamefont {Kresse}\ and\ \citenamefont
  {Furthm\"{u}ller}(1996)}]{Kresse_efficient_1996}%
  \BibitemOpen
  \bibfield  {author} {\bibinfo {author} {\bibfnamefont {G.}~\bibnamefont
  {Kresse}}\ and\ \bibinfo {author} {\bibfnamefont {J.}~\bibnamefont
  {Furthm\"{u}ller}},\ }\href {\doibase 10.1103/PhysRevB.54.11169} {\bibfield
  {journal} {\bibinfo  {journal} {Phys. Rev. B}\ }\textbf {\bibinfo {volume}
  {54}},\ \bibinfo {pages} {11169} (\bibinfo {year} {1996})}\BibitemShut
  {NoStop}%
\bibitem [{\citenamefont {Perdew}\ \emph {et~al.}(1996)\citenamefont {Perdew},
  \citenamefont {Burke},\ and\ \citenamefont
  {Ernzerhof}}]{Perdew_generalized_1996}%
  \BibitemOpen
  \bibfield  {author} {\bibinfo {author} {\bibfnamefont {J.~P.}\ \bibnamefont
  {Perdew}}, \bibinfo {author} {\bibfnamefont {K.}~\bibnamefont {Burke}}, \
  and\ \bibinfo {author} {\bibfnamefont {M.}~\bibnamefont {Ernzerhof}},\ }\href
  {\doibase 10.1103/PhysRevLett.77.3865} {\bibfield  {journal} {\bibinfo
  {journal} {Phys. Rev. Lett.}\ }\textbf {\bibinfo {volume} {77}},\ \bibinfo
  {pages} {3865} (\bibinfo {year} {1996})}\BibitemShut {NoStop}%
\bibitem [{\citenamefont {Sokolov}\ \emph {et~al.}(2004)\citenamefont
  {Sokolov}, \citenamefont {Skipidarov}, \citenamefont {Duvankov},\ and\
  \citenamefont {Shabunina}}]{Sokolov_chemical_2004}%
  \BibitemOpen
  \bibfield  {author} {\bibinfo {author} {\bibfnamefont {O.}~\bibnamefont
  {Sokolov}}, \bibinfo {author} {\bibfnamefont {S.}~\bibnamefont {Skipidarov}},
  \bibinfo {author} {\bibfnamefont {N.}~\bibnamefont {Duvankov}}, \ and\
  \bibinfo {author} {\bibfnamefont {G.}~\bibnamefont {Shabunina}},\ }\href
  {\doibase 10.1016/j.jcrysgro.2003.10.073} {\bibfield  {journal} {\bibinfo
  {journal} {J. Cryst. Growth}\ }\textbf {\bibinfo {volume} {262}},\ \bibinfo
  {pages} {442} (\bibinfo {year} {2004})}\BibitemShut {NoStop}%
\bibitem [{\citenamefont {Jnawali}\ \emph {et~al.}(2009)\citenamefont
  {Jnawali}, \citenamefont {Wagner}, \citenamefont {Hattab}, \citenamefont
  {Moller},\ and\ \citenamefont {Horn-von Hoegen}}]{jnawali_nucleation_2009}%
  \BibitemOpen
  \bibfield  {author} {\bibinfo {author} {\bibfnamefont {G.}~\bibnamefont
  {Jnawali}}, \bibinfo {author} {\bibfnamefont {T.}~\bibnamefont {Wagner}},
  \bibinfo {author} {\bibfnamefont {H.}~\bibnamefont {Hattab}}, \bibinfo
  {author} {\bibfnamefont {R.}~\bibnamefont {Moller}}, \ and\ \bibinfo {author}
  {\bibfnamefont {M.}~\bibnamefont {Horn-von Hoegen}},\ }\href {\doibase
  10.1103/PhysRevB.79.193306} {\bibfield  {journal} {\bibinfo  {journal} {Phys.
  Rev. B}\ }\textbf {\bibinfo {volume} {79}},\ \bibinfo {pages} {193306}
  (\bibinfo {year} {2009})}\BibitemShut {NoStop}%
\bibitem [{\citenamefont {Das}\ and\ \citenamefont
  {Rao}(2011)}]{Spin_das_2011}%
  \BibitemOpen
  \bibfield  {author} {\bibinfo {author} {\bibfnamefont {S.}~\bibnamefont
  {Das}}\ and\ \bibinfo {author} {\bibfnamefont {S.}~\bibnamefont {Rao}},\
  }\href {\doibase 10.1103/PhysRevLett.106.236403} {\bibfield  {journal}
  {\bibinfo  {journal} {Phys. Rev. Lett.}\ }\textbf {\bibinfo {volume} {106}},\
  \bibinfo {pages} {236403} (\bibinfo {year} {2011})}\BibitemShut {NoStop}%
\end{thebibliography}
\end{document}